\documentclass[%
 aip,
 amsmath,amssymb,
 reprint,%
]{revtex4-1}

\usepackage{graphicx}
\usepackage{dcolumn}
\usepackage{bm}

\usepackage[utf8]{inputenc}
\usepackage[T1]{fontenc}
\usepackage{mathptmx}
\usepackage{etoolbox}

\usepackage{color}
\usepackage{amsmath}
\usepackage{amssymb}
\usepackage{subfigure}
\usepackage{epsfig}
\usepackage{float}
\usepackage{lipsum}
\usepackage{cases}
\usepackage{appendix}
\usepackage{cancel}
\usepackage{soul}

\newcommand{\G}{\bm{G}}
\newcommand{\K}{\bm{K}}

\newcommand{\F}{\bm{F}}

\newcommand{\A}{\bm{A}}
\newcommand{\1}{\mathbb{I}}

\newcommand{\Ei}{\mathrm{Ei}}

\newcommand{\dif}{\mathop{}\!\mathrm{d}}

\newcommand{\vv}{\bm{v}} 

\newcommand{\rr}{\bm{r}}

\newcommand{\xxi}{\bm{\xi}}

\newcommand{\eeta}{\bm{\eta}}

\newcommand{\dbar}{\lower0.15ex\hbox{$\mathchar'26$}\mkern-12mu \dif}

\makeatletter
\def\@email#1#2{%
 \endgroup
 \patchcmd{\titleblock@produce}
  {\frontmatter@RRAPformat}
  {\frontmatter@RRAPformat{\produce@RRAP{*#1\href{mailto:#2}{#2}}}\frontmatter@RRAPformat}
  {}{}
}%

\makeatother
\begin{document}

\preprint{AIP/123-QED}

\title[]{Tailoring the escape rate of a Brownian particle by combining a vortex flow with a magnetic field}

\author{I. Abdoli}
\altaffiliation[Also at ]{Technische Universit\"at Dresden, Institut f\"ur Theoretische Physik, 01069 Dresden, Germany}
\affiliation{%
Leibniz-Institut  f\"ur Polymerforschung Dresden, Institut Theorie der Polymere, 01069 Dresden, Germany
}%
 \email{abdoli@ipfdd.de.} 
 
\author{H. L\"owen}%

\affiliation{ 
Institut f\"ur Theoretische Physik II: Weiche Materie, Heinrich-Heine-Universit\"at D\"usseldorf, D\"usseldorf, 40225, Germany
}%

\author{J.-U. Sommer}
 \altaffiliation[Also at ]{Technische Universit\"at Dresden, Institut f\"ur Theoretische Physik, 01069 Dresden, Germany}%
\affiliation{%
Leibniz-Institut  f\"ur Polymerforschung Dresden, Institut Theorie der Polymere, 01069 Dresden, Germany
}%
 
\author{A. Sharma}
 \altaffiliation[Also at ]{Technische Universit\"at Dresden, Institut f\"ur Theoretische Physik, 01069 Dresden, Germany}%
\affiliation{%
Leibniz-Institut  f\"ur Polymerforschung Dresden, Institut Theorie der Polymere, 01069 Dresden, Germany
}%
 

\begin{abstract}
The probability per unit time for a thermally activated Brownian particle to escape over a potential well is in general well-described by Kramers theory. Kramers showed that the escape time decreases exponentially with increasing barrier height.  The dynamics slow down when the particle is charged and subjected to a Lorentz force due to an external magnetic field. This is evident via a rescaling of the diffusion coefficient entering as a prefactor in the Kramers escape rate without any impact on the barrier-height-dependent exponent. Here we show that the barrier height can be effectively changed when the charged particle is subjected to an external vortex flow. While the external vortex alone does not affect the mean escape time of the particle, when combined with a magnetic field it effectively pushes the fluctuating particle either radially outside or inside depending on its sign relative to that of the magnetic field.  In particular, the effective potential over which the particle escapes can be changed to a flat, a stable, and an unstable  potential by tuning the signs and magnitudes of the external vortex and the applied magnetic field. Notably, the last case corresponds to enhanced escape dynamics. 

\end{abstract}

\maketitle

\section{\label{sec:level1}Introduction}
A Brownian particle undergoes erratic motion as a result of its collisions with the solvent molecules. If the particle is being initially put at the bottom of a potential well, the thermal activation of the particle may cause an escape from the potential well over an energetic barrier.  Using the flux-over-population method~\cite{farkas1927keimbildungsgeschwindigkeit}, Kramers first derived the escape rate of a Brownian particle over an energy barrier moving in a bistable potential, regardless of what happens after this escape~\cite{kramers1940brownian}. He showed that the probability per unit time for the particle to escape the potential well exponentially decays with the height of the energy barrier. Kramers derived limiting expressions for weak friction and strong damping and realized a global maximum at some intermediate value of the damping, which is known as \textit{Kramers turnover}~\cite{grabert1988effect, mccann1999thermally,rondin2017direct}. The problem has been generalized to include memory friction~\cite{grote1980stable, carmeli1982non, ianconescu2015study} and athermal fluctuations~\cite{hanggi1984bistable, jung1988bistability, sharma2017escape, scacchi2019escape,caprini2019active} and was extended to quantum field theory~\cite{berera2019formulating, darme2019generalized}.

While Kramers’ framework and its extensions have been thoroughly studied with the relevant deterministic potential force fields~\cite{fleming1993activated,talkner1995new, hanggi1982thermally, pollak1986theory, pollak1989theory,pollak2013improvements}, much less is known when the deterministic force is nonconservative, namely when it is not of potential type~\cite{baura2013tuning, mondal2018generic, filliger2007kramers}. Recently, by taking into account a nonconservative force in the form of Lorentz force, we have studied the escape dynamics of a two-dimensional Brownian system with broken spatial symmetry via two noises with different strengths~\cite{abdoli2022escape}. We have shown that while the escape process becomes anisotropic (i.e. particles tend to escape the potential well more along the axis with larger noise strength) due to two different noises, when subjected to an external magnetic field, the spatial symmetry can be restored~\cite{abdoli2022escape}.  However, to our knowledge, it is expected that the escape process is reduced (or unaffected in the direction of the applied magnetic field) by external constant magnetic fields~\cite{filliger2007kramers, abdoli2022escape} which is evident via a rescaling of the diffusion coefficient. It has been shown that the combined influence of a nonconservative force and a magnetic field may cause an instability in the system~\cite{lee2019nonequilibrium}. Here, taking advantage of such an instability, we show that the Lorentz force due to a constant magnetic field can result in enhanced escape dynamics.

In this work, we study the escape dynamics of a Brownian particle from a harmonic trap which is cut-off at a certain distance in the presence of an external vortex and  the Lorentz force due to an external constant magnetic field. Taking advantage of the spatial isotropy in the system we derive an exact expression for the mean first passage time. While the external vortex alone does not affect the escape dynamics, we observe a nontrivial result when an external magnetic field is present: the mean first passage time can be reduced or enhanced. This is attributed to the shape change of the effective potential well.  By tuning the external magnetic field or alternatively the strength of the external vortex the effective potential can change shape to a flat, a stable, or an unstable potential. This means that by tuning either parameters the barrier energy over which the particle may escape can be effectively altered to a smaller or larger one whose origin can be understood as follows: the combination of the vortex flow and the magnetic field effectively pushes the fluctuating particle either radially outside or inside depending on their signs. In other words, the combination of the two fields, which individually induces no radial force, gives rise to a radial force.
In what follows, we first introduce the model. Next we calculate the mean first passage time, which can be written in terms of an effective potential. We then study the trends of the escape time with respect to the magnetic field strength and the vortex flow and finally we discuss several experimental realizations of the set-up considered here.

\section{Model}
\begin{figure}
\includegraphics[width=0.7\linewidth]{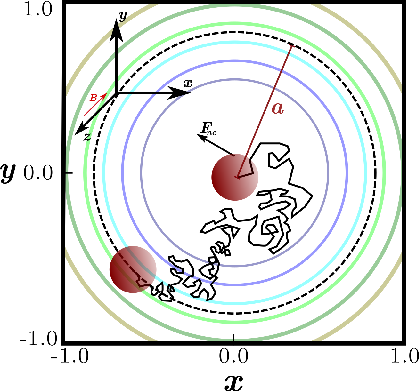}
\caption{\label{fig01}A single charged particle diffusing in a two-dimensional harmonic potential $U(x, y)=k(x^2+y^2)/2$, shown by concentric contours, with $k$ being its stiffness. The particle is subjected to an external magnetic field $B$ in the $-\hat{z}$ direction and a nonconservative force $\F_{nc}=\epsilon(-y, x)$ with $\epsilon$ being its strength. The nonconservative force is shown for $\epsilon>0$. The particle can escape the trap when reaches the boundary, truncated at $r=a$, shown by dashed circle, where $r=\sqrt{x^2+y^2}$ is the distance from the origin.}
\end{figure}

We consider an overdamped charged Brownian particle with the charge $q$ subjected to an external magnetic field $B$ in the $-\hat{z}$ direction. Since the Lorentz force due to the field does not affect the motion of the particle in the $z$ direction, we effectively reduce the system to a two-dimensional one and study the motion of the particle in the $xy$ plane. The particle is trapped in an isotropic potential $U(x, y)=k(x^2+y^2)/2$ and undergoes a vortex flow due to the nonconservative force $\F_{nc}=\epsilon(-y, x)^\top$. Here $k$ and $\epsilon$ are the stiffness of the potential and the strength of the nonconservative force, respectively. A schematic of the system is shown in Fig.~\ref{fig01}. It is experimentally and theoretically known that even statically optically trapped Brownian particles in the overdamped limit represent nonequilibrium behavior characterized by Brownian vortices. This is due to the nonconservative forces generated by optical scattering forces~\cite{sun2010minimal, sun2009brownian, roichman2008influence, moyses2015perturbative}.  Moreover,  by applying a prescribed external vortex flow field such as a rotating bucket to an underdamped Brownian particle one can induce similar terms  to the nonconservative force, i.e. $-\epsilon y$ and $\epsilon x$~\cite{liebchen2019optimal}.

It has been shown that  the overdamped dynamics of the particle derived by simply setting the inertia term to zero  can yield an incorrect description in the presence of a magnetic field~\cite{vuijk2019anomalous}. In this case, the overdamped Langevin equation describing dynamics of the system can be derived using the low-mass approach~\cite{chun2018emergence, abdoli2022tunable, abdoli2022escape}, which can be written as


\begin{eqnarray}
\dot{x} & = & \frac{1}{\gamma(1+\kappa^2)}\left[-kx-\epsilon y+k\kappa y-\epsilon\kappa x\right] + \xi_x(t), \label{eq01} \\
\dot{y} & = & \frac{1}{\gamma(1+\kappa^2)}\left[-ky+\epsilon x-k\kappa x-\epsilon\kappa y\right] + \xi_y(t), \label{eq02}
\end{eqnarray}
where $\gamma$ is the friction coefficient and $\kappa=qB/\gamma$ is the diffusive Hall parameter quantifying the strength of the Lorentz force relative to the frictional force. We note that $\kappa$ can be positive or negative depending on the sign of the applied magnetic field. Here $\xxi(t)=(\xi_x, \xi_y)^\top$ is Gaussian nonwhite noise with zero mean and time correlation $\langle\xxi(t)\xxi^\top(t')\rangle  = T\G^{-1}\delta_+(t-t') + T(\G^{-1})^\top\delta_-(t-t')$ where $T$ is the temperature, $\G = \gamma\bigl( \begin{smallmatrix}1 & \kappa\\ -\kappa & 1\end{smallmatrix}\bigr)$, and the notations $\delta_{\pm}(s=t-t')$ are the modified Dirac delta functions which are zero for $s\neq 0$ while $\int_0^\infty \dif s\delta_+(s)=\int_{-\infty}^0\dif s\delta_-(s)=1$ and $\int_0^\infty \dif s\delta_-(s)=\int_{-\infty}^0\dif s\delta_+(s)=0$. Throughout this work we set the Boltzmann constant $k_B$ to unity. Length and time are measured in units of $\sqrt{T/k}$ and $\gamma/k$, respectively.

We use It\^o calculus to reduce the Langevin equations in Eq.~\eqref{eq01} and Eq.~\eqref{eq02} to a one-dimensional problem for the variable $r=\sqrt{x^2+y^2}$, which is given as~\cite{gardiner2009stochastic} 
\begin{equation}
\label{eq03}
\dif r= \frac{1}{1+\kappa^2}\left[-\frac{k+\epsilon\kappa}{\gamma}r+\frac{D}{r}\right]\dif t +\sqrt{\frac{2D}{1+\kappa^2}}\eta(t)\dif t,
\end{equation}
where $D=T/\gamma$ is the coefficient of a freely diffusing particle and $\eta(t)$ is Gaussian white noise with zero mean and the Dirac delta time correlation $\langle\eta(t)\eta(t')\rangle=\delta(t-t')$. The terms in the square brackets on the right hand side of Eq.\eqref{eq03} describe the force on the the particle due to an effective potential, given as
\begin{figure}
\includegraphics[width=1\linewidth]{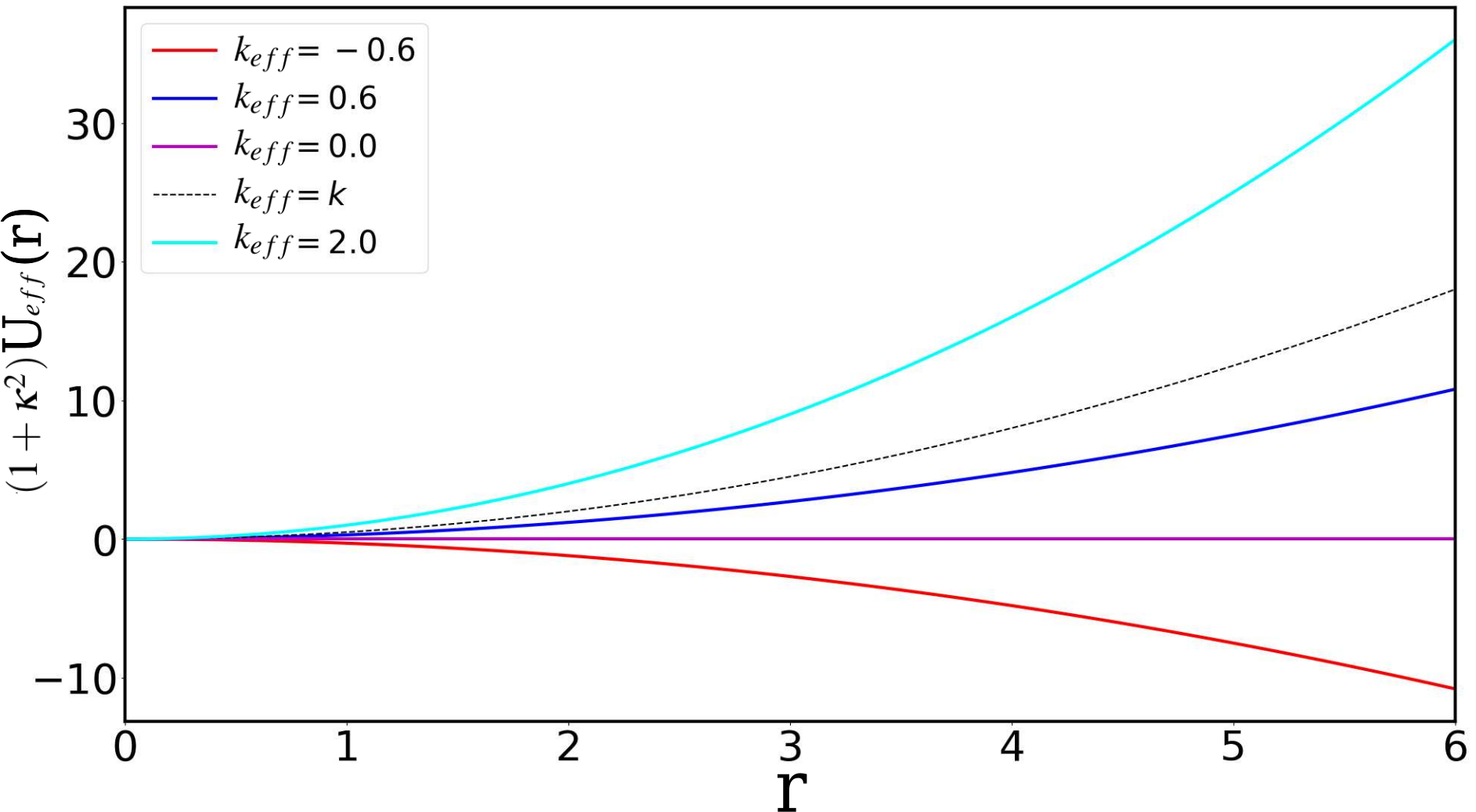}
\caption{\label{fig02}The effective potential from Eq.~\eqref{eq04} for different values of the stiffness $k_{eff} = k+\epsilon\kappa$. By varying the parameter $\kappa$ (or $\epsilon$) the effective potential can change shape to a stable one if $k_{eff}>0$, a flat one if $k_{eff}=0$, or to an unstable one if $k_{eff}<0$. }
\end{figure}

\begin{equation}
U_{eff}(r) = \frac{k_{eff}}{2\gamma(1+\kappa^2)}r^2-\frac{D}{1+\kappa^2}\log(r),
\label{eq04}
\end{equation}
where $k_{eff}=k+\epsilon\kappa$ is the stiffness of the effective potential. As it is evident from the effective stiffness, in the absence of the vortex flow the potential simply gets rescaled by the factor $1/(1+\kappa^2)$, as we have shown in the supplemental information of Ref.~\cite{abdoli2022escape}. Moreover, in the absence of the magnetic field there is no effect of the vortex flow on the effective potential and Eq.~\eqref{eq04} reduces to the well known results in Ref.\cite{gardiner2009stochastic} for a rotationally symmetric Ornstein-Uhlenbeck process in two dimensions.
The second term on the right and side comes from the transformation to $r$ and corresponds to an extremely repulsive potential at the origin due to reduced number of states on the circle of radius $r$. This term influences the motion of the particle only near the origin and is negligible for larger distances as compared to the first term. 

Figure.~\ref{fig02} represents the scaled effective potential from Eq.~\eqref{eq04} for different values of the parameter $k_{eff}$ without the logarithmic term. By tuning the diffusive Hall parameter or alternatively the strength of the nonconservative force, the effective potential changes shape: the potential is stable if $k_{eff}>0$, flat if $k_{eff}=0$, and unstable if $k_{eff}<0$. It becomes simple quadratic potential in the absence of $\epsilon$ or/and $\kappa$.

\section{Mean Escape Time}
\begin{figure}
\includegraphics[width=1\linewidth]{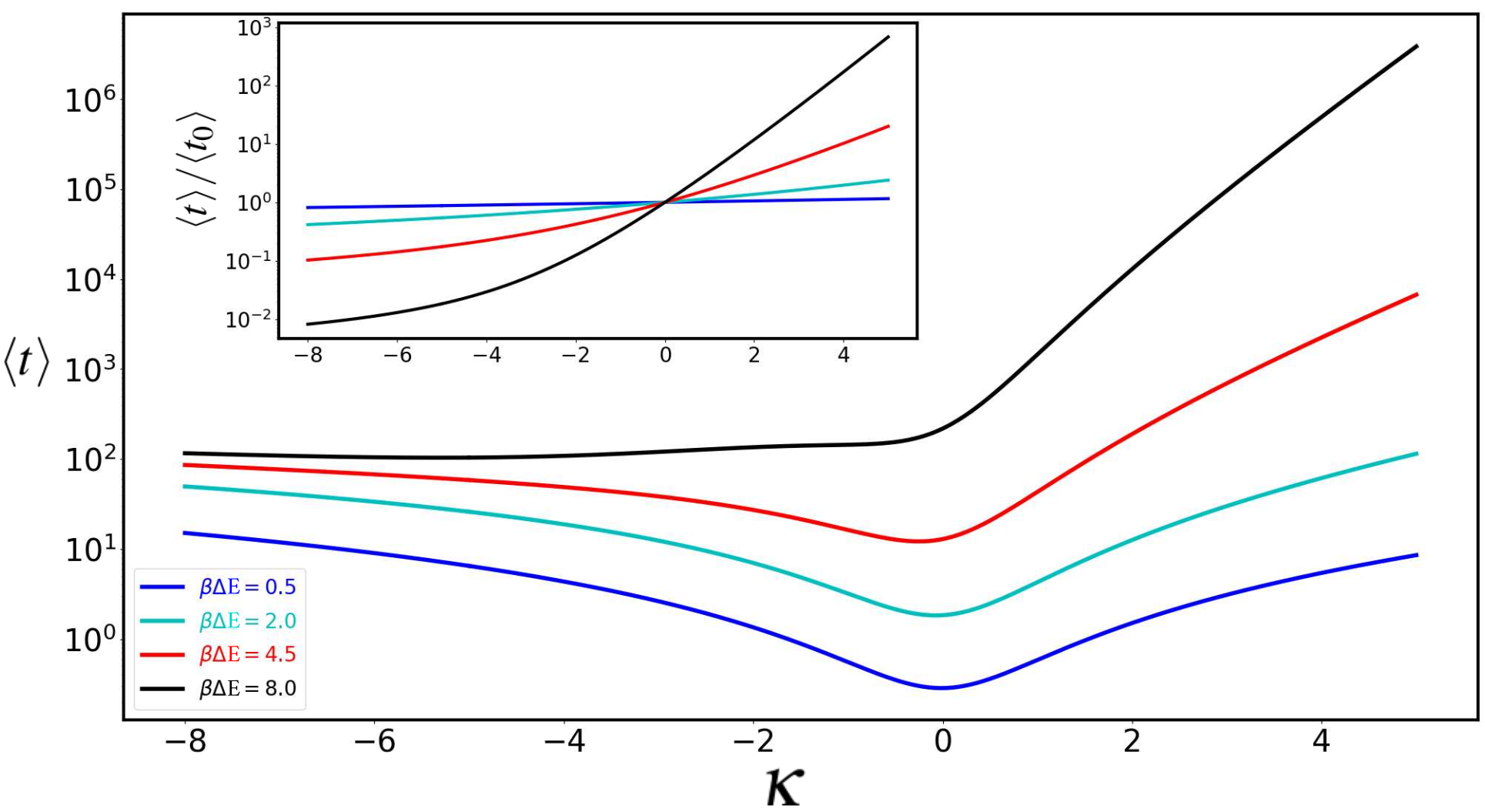}
\caption{\label{fig03}The mean escape time as a function of the diffusive Hall parameter $\kappa$ from Eq.~\eqref{MFPTp} and Eq.~\eqref{MFPT0} for different values of the scaled barrier height $\beta\Delta E$ with $\beta=1.0$, $\gamma=1.0$ and $\epsilon=0.2$. Obviously the mean escape time increases with increasing the barrier height. It can increase or decrease by tuning the parameter $\kappa$: the presence of an external vortex field can work together with the applied magnetic field to  effectively push the fluctuating particle either radially outside, if $\kappa<-k/\epsilon$, or inside, if $\kappa>-k/\epsilon$. The former corresponds to the case in which the combination helps the particle to escape. The point $\kappa=0$ corresponds to unaffected escape time by the external vortex (i.e. $k_{eff}=k$). In the inset, we show the mean escape time which is scaled by the mean escape time in the absence of the external vortex $\langle t_0\rangle$ where the subscript $0$ indicates zero strength length of the vortex flow.  It implies that the mean escape time can decrease with increasing $\kappa$ as compared to the mean escape time without the vortex flow. }
\end{figure}

We consider a particle which is trapped in an isotropic potential $U(x, y)$ which taking advantage of the spatial symmetry whose distance from the origin, $r=|\rr|$, can be described by Eq.~\eqref{eq03}. We are interested in the mean time at which the particle reaches the boundary, truncated at $r=a$, as shown in Fig.~\ref{fig01}.  As we show in the Appendix \ref{appendix}, the mean escape time can be exactly calculated from Eq.~\eqref{eq03} which reads 

\begin{equation}
 \langle t\rangle= \frac{\gamma(1+\kappa^2)}{2k_{eff}}\left[\mathrm{Ei}\left(\beta\Delta E_{eff}\right)-\log\left(\beta\Delta E_{eff}\right)-\gamma_{EM}\right],
\label{MFPTp}
\end{equation}
if $ k_{eff} > 0$ corresponding to the effective stable potential  and 
\begin{equation}
 \langle t\rangle= \frac{\gamma(1+\kappa^2)}{2k_{eff}}\left[-\mathrm{Ei}\left(-\beta|\Delta E_{eff}|\right)+\log\left(\beta|\Delta E_{eff}|\right)+\gamma_{EM}\right],
\label{MFPTn}
\end{equation}
if $ k_{eff} < 0$ corresponding to the unstable effective potential
where $\beta$ is the inverse of the temperature, $\gamma_{EM}$ is the Euler-Mascheroni constant, and $\Ei(x)$ is the exponential integral. Here $\Delta E_{eff}=\Delta E + \epsilon\kappa a^2/2$ is the effective barrier energy which is the real barrier height $\Delta E=ka^2/2$ augmented by  the coupling between the magnitude of the applied magnetic field and the strength of the external vortex.  Using the series expansion of the exponential integral at $k_{eff}=0$ for Eq.~\eqref{MFPTp} and Eq.~\eqref{MFPTn}, the mean escape time  for the effective flat potential reads 
\begin{equation}
\langle t\rangle\sim\frac{(1+\kappa^2)}{4D}a^2,
\label{MFPT0}
\end{equation}
which is the mean escape time for a freely diffusing particle scaled by $1+\kappa^2$. In the limit of large barrier heights the exponential integral in Eq.~\eqref{MFPTp} can be expanded and as a consequence the mean escape time reduces to $\langle t\rangle\sim\gamma(1+\kappa^2)\exp(\beta\Delta E_{eff})/(2k_{eff}\beta\Delta E_{eff})$. In the absence of the external vortex, which corresponds to $\epsilon=0$,  the result reduces to the Kramers result  rescaled by $1+\kappa^2$ arising from the trivial rescaling of the diffusion coefficient. The expression becomes the same as the Kramers one when the magnetic field is absent $\kappa=0$. This confirms that the external vortex field alone does not affect the mean escape time. The intuitive reason for that is that, for $\kappa=0$, the presence of a vortex field only changes the azimuthal motion but not the radial one which leaves the redial particle escape unaffected.

\begin{figure}
\includegraphics[width=1\linewidth]{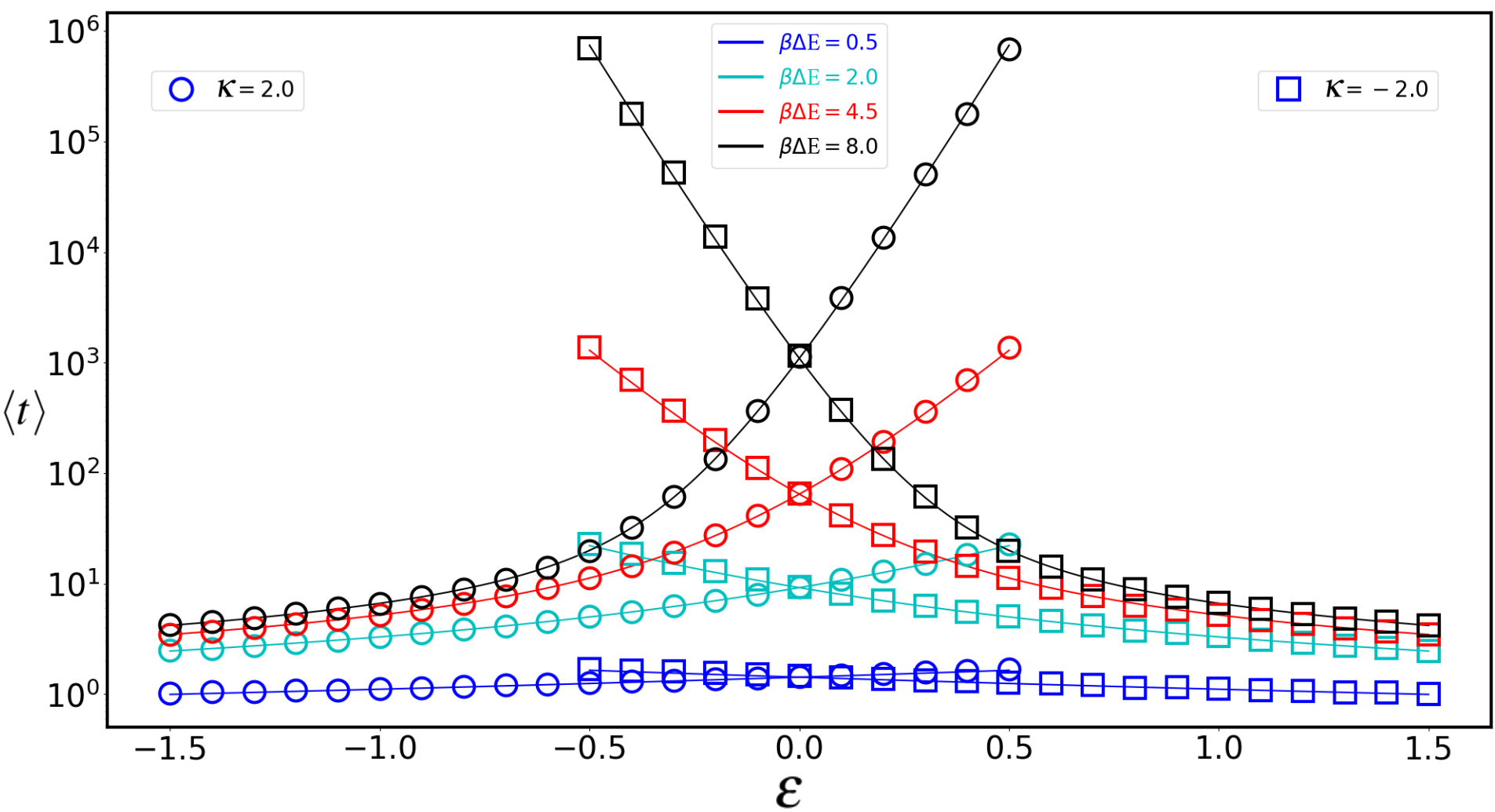}
\caption{\label{fig04}  The mean escape time with respect to the strength of the conservative force $\epsilon$ from Eq.~\eqref{MFPTp} and Eq.~\eqref{MFPT0} for different values of the scaled barrier heights with $\beta=1.0$ and $\gamma=1.0$.  The lines with circles and squares correspond to the results with $\kappa=2.0$ and $\kappa=-2.0$, respectively. The mean escape time can increase or decrease with increasing the strength of the vortex flow, which  depends on its sign relative to that of $\kappa$ and their magnitude compared to the stiffness of the potential $k$.}
\end{figure}


Figure~\ref{fig03} shows the mean escape time with respect to the diffusive Hall parameter $\kappa$. Obviously it takes the particle longer time to escape over larger barrier heights as is evident in the figure. The magnetic field together with the vortex flow creates additional fluctuations in radial direction which can be directed either outwards or inwards depending on its sign. The former corresponds to the case in which the combination of the vortex flow and the magnetic field helps the particle to escape.
The inset shows the mean escape time scaled by the mean escape time in the absence of the external vortex, which is indicated by the subscript $0$. The mean escape time can decrease with increasing  magnetic field as compared to the mean escape time without the vortex flow and remains almost constant for small barrier height. 

In Fig.\ref{fig04}, we show that tuning the strength of the vortex flow is an alternative way to vary the mean escape time which is evident in Eq.\eqref{MFPTp} and Eq.\eqref{MFPTn} via the production of the two parameters, i.e. $\epsilon\kappa$. Therefore the similar trends are expected.  The figure represents the mean escape time with respect to the parameter $\epsilon$ for a system with $\kappa=2.0$, denoted by lines with circles, and a system with $\kappa=-2.0$, denoted by lines with squares. Our results imply that the mean escape time can be decreased or increased by tuning the vortex flow strength depending on its sign relative to that of the magnetic field and their magnitude compared to the stiffness of the potential $k$. 


\section{Discussion}
 In this work, we studied the effect of a vortex flow  on the escape dynamics of a Brownian magneto-system made of single charged Brownian particle subjected to an external magnetic field.  We expressed the potential in an effective form which can change shape to a stable, a flat, or an unstable potential depending on the stiffness of the effective potential. Taking advantage of the spatial isotropy in the system we obtained an exact expression for the mean escape time.   In the absence of the external vortex, exerted by the  nonconservative force, the Lorentz force due to the external magnetic field slows down the dynamics of the system without any qualitative change, which is evident via the trivial rescaling of the diffusion coefficient. We showed that while the external vortex alone does not affect the mean escape time, when coupled to the magnetic field it can enhance or reduce the escape time: this is intuitive as the magnetic field together with the vortex flow creates additional fluctuations in radial direction which can be directed either outwards or inwards depending on its sign. In other words, the combination of the two fields, which individually induces no radial force, gives rise to a radial force. We showed that the barrier over which the particle escapes can be effectively changed to a larger or smaller one depending on the relative signs of the strength of the vortex flow and the applied magnetic field as well as their magnitude compared to the stiffness of the potential in which the particle is trapped.  Moreover, the trap can be effectively switched-off by an appropriate sign and value of the magnetic field. 
 


A possible experimental realisation is to trap the particle using optical tweezers either in a radio-frequency plasma sheath with a vertical magnetic field~\cite{carstensen2009effect, piel2017plasma} or in a rotating frame of reference. By rotating the reference frame a Coriolis force can be induced which acts  the same as the Lorentz force due to an external magnetic field~\cite{kahlert2012magnetizing, hartmann2013magnetoplasmons, hartmann2019self}. As it has been shown that even statically optically trapped Brownian particles undergoe a nonconservative force induced by optical scattering forces~\cite{sun2010minimal, sun2009brownian, roichman2008influence, moyses2015perturbative, mangeat2019role}, we expect that the study of the enhanced escape dynamics does not require an additional external vortex. Another possibility is to apply a rotating bucket to an underdamped Brownian particle which induces similar terms to the nonconservative force in the overdamped limit~\cite{liebchen2019optimal}.

From a future perspective, it could be interesting to study the escape dynamics of an opposite charged dimer~\cite{shinde2022strongly}. In the limit of low persistence length, an active chiral particle follows curved trajectories, similar to the Brownian motion of a charged particle~\cite{van2008dynamics, scholz2021surfactants} under a magnetic field. Therefore, another study of interest would be the escape dynamics of a chiral active Brownian particle in the presence of an external vortex. Finally, it could be interesting to study how an external magnetic field can affect an active turnover for an active particle in a bistable potential~\cite{militaru2021escape}--an optimal correlation time where  the  transition  rate  is  maximized-- and how an external vortex influences new turnovers observed in the presence of a fluctuating magnetic field~\cite{baura2013tuning, mondal2018generic}.


\begin{acknowledgments}
A. Sharma and H. L\"owen acknowledge the support by the Deutsche Forschungsgemeinschaft (DFG) within the projects SH 1275/3-1 (A.S.) and LO 418/25-1 (H.L.).
\end{acknowledgments}

\section*{Data Availability Statement}

The data that support the findings of this study are available from the corresponding author upon reasonable request.

\section*{Author Declarations}
The authors have no conflicts to disclose.

\appendix

\section{Derivation of the mean escape time}
\label{appendix}
The main purpose of this section is to derive the mean escape time in Eq.~\eqref{MFPTp} to Eq.~\eqref{MFPT0}. We start with the underdamped Langevin equation describing the dynamics of a charged Brownian particle with mass $m$ and charge $q$ subjected to a magnetic field $B$ in the $-\hat{z}$ direction. The velocity Langevin equation for the position $\rr=(x, y)^\top$ and the velocity $\vv=(v_x, v_y)^\top$ of the particle under the effect of the linear nonconservative force $\F_{nc}=\epsilon(-y, x)^\top$ and the conservative force $\F_c=-k(x, y)^\top$ due to the isotropic potential $U(x, y)=k(x^2+y^2)/2$, can be written as
\begin{equation}
\label{underdampedlangevin}
m\dot{\vv} = -\K\rr-\G\vv(t)+\sqrt{2\gamma T}\eeta(t),
\end{equation}
where  $\eeta(t)=(\eta_x(t), \eta_y(t))^\top$ is the Gaussian white noise with zero mean and Dirac delta correlation $\langle\eeta(t)\eeta^\top(t')\rangle=\delta(t-t')$ with $\gamma$ being the friction coefficient and $T$ the temperature. The matrices $\G$ and $\K$ are defined as
\begin{equation}
\label{matrixG}
\G = \gamma\left( \begin{array}{cc}
1 & \kappa \\
-\kappa & 1 \\
\end{array}\right),  \,\,\,\,\,  \K = \left( \begin{array}{cc}
k & \epsilon \\
-\epsilon & k \\
\end{array}\right),
\end{equation}
with $\kappa=qB/\gamma$ being the diffusive Hall parameter which quantifies the strength of the Lorentz force relative to the frictional force. Using the low-mass approach, the corresponding overdamped Langevin equation can be written as~\cite{chun2018emergence, abdoli2022tunable, abdoli2022escape}

\begin{equation}
\dot{\rr} = \A \rr + \xxi(t),
\label{lowmasslimit}
\end{equation}
where $\A=\G^{-1}\K$ and $\xxi(t)=(\xi_x, \xi_y)^\top$ is Gaussian nonwhite noise with 

\begin{eqnarray}
\langle\xxi(t)\rangle & = & 0,  \label{zeromeanmean} \\
 \langle\xxi(t)\xxi^\top(t')\rangle & = & T\G^{-1}\delta_+(t-t') + T(\G^{-1})^\top\delta_-(t-t'), \,\,\,\,\,\,\,\,\,\,\,\,\,\,\,\, \label{timecorrelation}
\end{eqnarray}
where  $\delta_{\pm}(s=t-t')$ are the modified Dirac delta functions which are zero for $s\neq 0$ while $\int_0^\infty \dif s\delta_+(s)=\int_{-\infty}^0\dif s\delta_-(s)=1$ and $\int_0^\infty \dif s\delta_-(s)=\int_{-\infty}^0\dif s\delta_+(s)=0$. 

Equation~\eqref{lowmasslimit} can be rewritten as Eq.~\eqref{eq01} and Eq.~\eqref{eq02} in the Cartesian coordinates and thereafter using It\^o calculus can be reduced to a one-dimensional equation for the variable $r$, which is the distance from the origin and is given by Eq.~\eqref{eq03}. The mean time for the particle to escape the trap, truncated at $r=a$, can be obtained by the following equation
\begin{equation}
\langle t\rangle = \frac{1+\kappa^2}{D}\int_0^{a}y^{-1} \exp\left[\frac{k_{eff}}{2\gamma D}y^2\right]dy \int_0^y z\exp\left[-\frac{k_{eff}}{2\gamma D}z^2\right]dz,
\label{MFPT01}
\end{equation} 
where $D=T/\gamma$ is the diffusion coefficient for a freely moving particle. This equation can be exactly solved: using a change of variables the second integral on the right hand side gives $(\gamma D/k_{eff})\left[1-\exp\left(-k_{eff}y^2/2\gamma D\right)\right]$. By substitution of this solution into Eq.~\eqref{MFPT01}, the resulting integral can be exactly solved which gives  Eq.~\eqref{MFPTp} and Eq.~\eqref{MFPTn}.

\section*{References}

\providecommand{\noopsort}[1]{}\providecommand{\singleletter}[1]{#1}%

\end{document}